\documentstyle[aps,12pt]{revtex}

\begin{document}
\author{DONG MI\thanks{%
E-mail: dongmi@dlut.edu.cn}, HE-SHAN SONG\thanks{%
E-mail: hssong@dlut.edu.cn}}
\address{Department of Physics, Dalian University of Technology, Dalian 116024,\\
P.R.China}
\author{YING\ AN}
\address{Institute of Theoretical Physics, Academia Sinica, P.O.\ Box 2735, Beijing\\
100080, P. R. China}
\title{A PHYSICAL EXPLANATION FOR THE TILDE SYSTEM IN THERMO FIELD DYNAMICS}
\maketitle

\begin{abstract}
For a two-body quantum system, any pure state can be represented by a
biorthogonal expression by means of Schmidt decomposition. Using this in the
composite system which include a thermodynamic system and its surroundings,
it is found that the tilde system in thermo field dynamics is just the
surroundings of the real system.
\end{abstract}

\vskip 0.5cm


\baselineskip 24pt

~In 1975, Y. Takahashi and H. Umezawa constructed a quantum field theory at
finite temperature: Thermo Field Dynamics (TFD)$^1$. In this theory, the
``fictitious'' (or tilde) system is essential to convert thermal statistical
average into expectation value with respect to a pure state. The central
idea of TFD is the doubling of the Hilbert space of states. The operators on
this doubled space are effectively doubled as well$^{1,2}$. After that, most
of the techniques in quantum field theory can be effectively used to deal
with the problems in statistical physics. However, so far there exists a
question in TFD which has not been answered satisfactorily: what is the
physical meaning of the tilde system? Is it really a fictitious system? The
discussion of the physical meaning for the tilde system started from the
birth of TFD. For example, Takahashi and Umezawa suggested in their original
paper that the particles in the tilde system may be regarded as the holes$^1$%
. Y. X. Gui has suggested it may be regarded as the mirror world of our real
one$^3$. Before the passing away of Umezawa, he still referred to the tilde
system and tried to explain its physical meaning. He pointed out that the
tilde operator may be called the thermal operator for all the thermal
quantities need the tilde operators$^4$. We think these explanations are
ambiguous and should belong to guesswork without good ground. In this
letter, we first consider a two-body quantum system, its any pure state can
be represented by a biorthogonal expression by means of Schmidt
decomposition. Notice that any thermodynamic system is not a truly isolated
system, it is always in contact with the surroundings. Using the idea about
Schmidt decomposition in the composite isolated system which include a
thermodynamic system and its surroundings, we find that the tilde system in
thermo field dynamics is just the surroundings of the real system.

Firstly, let us imagine an isolated system with an definite state vector$%
\left| \psi \right\rangle $. This isolated system will be supposed a
composite system and to consist of two parts. One is the system whose
properties we wish to study. The remainder, to be called the surroundings of
the system, will not be observed. The system to be measured has a complete
set of commuting observables whose eigenvalues will be labeled collectively
as a Roman letter, as $i$. The surroundings have a complete set of commuting
observables whose eigenvalues will be labeled collectively as a Greek
letter, as $\mu $. The observables belonging to the system and surroundings
commute because they refer to different things, so a basis for the isolated
system, system (denoted by $A$) plus surroundings (denoted by $B$), is the
set of simultaneous eigenstates$^5$ $\left| i\right\rangle \otimes \left|
\mu \right\rangle \equiv \left| i,\mu \right\rangle $. For simplicity, it
will be assumed that all the eigenvalues are discrete, so the completeness
relation is

\begin{equation}
\sum\limits_{i,\mu }\left| i,\mu \right\rangle \left\langle i,\mu \right| =%
\widehat{I}_{AB}.
\end{equation}
By using equation (1), $\left| \psi \right\rangle $ can be expanded as

\begin{equation}
\left| \psi \right\rangle =\sum\limits_{i,\mu }a_{i,\mu }\left| i,\mu
\right\rangle ,
\end{equation}
with the expansion coefficients $a_{i,\mu }=\left\langle i,\mu \right|
\left| \psi \right\rangle .$

An observable $\widehat{F}_A$ which refers to system $A$ alone, when
considered as an observable of the composite system, should have the
following form

\begin{equation}
\widehat{F}=\widehat{F}_A\otimes \widehat{I}_B,
\end{equation}
where $\widehat{I}_B$ is identity operator which acts on the Hilbert space
of the surroundings. The expectation value of $\ \widehat{F}_A\otimes 
\widehat{I}_B$ in the state $\left| \psi \right\rangle $ is as usual

\begin{equation}
\left\langle \widehat{F}_A\otimes \widehat{I}_B\right\rangle =\left\langle
\psi \left| \widehat{F}_A\otimes \widehat{I}_B\right| \psi \right\rangle .
\end{equation}
With equation (2), we can write this as

\begin{eqnarray}
\left\langle \widehat{F}_A\otimes \widehat{I}_B\right\rangle
&=&\sum\limits_{j,\nu }a_{j,\nu }^{*}\left\langle j,\nu \right| \left| 
\widehat{F}_A\otimes \widehat{I}_B\right| \sum\limits_{i,\mu }a_{i,\mu
}\left| i,\mu \right\rangle  \nonumber \\
&=&\sum\limits_{i,j,\mu }a_{j,\mu }^{*}a_{i,\mu }\left\langle j\right| 
\widehat{F}_A\left| i\right\rangle  \nonumber \\
&=&tr_A\left( \widehat{\rho }_A\widehat{F}_A\right) ,
\end{eqnarray}
where

\begin{eqnarray}
\widehat{\rho }_A &=&\sum\limits_{i,j,\mu }a_{i,\mu }a_{j,\mu }^{*}\left|
i\right\rangle \left\langle j\right|  \nonumber \\
&=&tr_B\left( \widehat{\rho }_{AB}\right)
\end{eqnarray}
is called the reduced density operator (or matrix) of the system$^6$.

It is easy to see that $\widehat{\rho }_A$ has the following properties

\begin{equation}
\widehat{\rho }_A\dagger =\widehat{\rho }_A,
\end{equation}

\begin{equation}
tr_A\widehat{\rho }_A=1,
\end{equation}

\begin{equation}
\widehat{\rho }_A\text{ is semi-positive definite}.
\end{equation}
Since the density matrix $\widehat{\rho }_A$ is a semi-positive definite
Hermitian matrix, it can be diagonalized and it will be have non-negative
real eigenvalues. Notice that the relation $\widehat{\rho }_A^2=\widehat{%
\rho }_A$ , in general, does not hold. This means that the system is in a
mixed state. Let the set $\{|i>\}$ be the complete orthonormal eigenstates
of the density matrix $\widehat{\rho }_A$ and $\{P_i\}$ its eigenvalues ($%
P_i\geq 0$ since $\widehat{\rho }_A$ is semi-positive definite), then the
density matrix can be written in the form

\begin{equation}
\widehat{\rho }_A=\sum\limits_iP_i\left| i\right\rangle \left\langle
i\right| .
\end{equation}
where $P_i$ is the probability of finding the system in state $\left|
i\right\rangle $ and $\sum\limits_iP_i=1$. In this way, the pure state $%
\left| \psi \right\rangle $ of the composite system can be rewritten as the
following form

\begin{equation}
\left| \psi \right\rangle =\sum\limits_iP_i^{1/2}\left| i,\alpha
_i\right\rangle ,
\end{equation}
where

\begin{equation}
\left| \alpha _i\right\rangle =\sum\limits_\mu P_i^{-1/2}a_{i,\mu }\left|
\mu \right\rangle .
\end{equation}
Here, $\{|\alpha _i>\}$ is a set of complete orthonormal states of the
surroundings. In fact, equation (11) is the Schmidt decomposition of a pure
state of the composite system$^7$, and the set $\{|i,\alpha _i>\}$ is called
Schmidt basis. From the viewpoint of the surroundings $B$, a pure state of
the composite system ($A$+$B$) appears as a mixed state of the system $A$,
described by a density matrix obtained by tracing over the degrees of
freedom of the surroundings, and the density matrix $\widehat{\rho }_A$ is
diagonal in the Schmidt basis. That is equation (10) can also be obtained by
the following way

\begin{eqnarray}
\widehat{\rho }_A &=&tr_B\left( \widehat{\rho }_{AB}\right) =tr_B\left(
\left| \psi \right\rangle \left\langle \psi \right| \right)  \nonumber \\
&=&\sum\limits_{i,j,k}P_i^{1/2}P_j^{1/2}\left\langle \alpha _k\right| \left|
i,\alpha _i\right\rangle \left\langle j,\alpha _j\right| \left| \alpha
_k\right\rangle  \nonumber \\
&=&\sum\limits_iP_i\left| i\right\rangle \left\langle i\right| .
\end{eqnarray}

We denote the expectation value of a quantity in mixed state by the symbol $%
\overline{\left\langle {}\right\rangle }$, for this expectation value is a
double averages including the thermo and quantum averages. Thus we can
express equation (5) as

\begin{equation}
\left\langle \widehat{F}_A\otimes \widehat{I}_B\right\rangle =\overline{%
\left\langle \widehat{F}_A\right\rangle }.
\end{equation}
Using equation (11) and (10) or (13), the above equation can be written in
the Schmidt basis as

\begin{equation}
\sum\limits_{i,j}P_i^{1/2}P_j^{1/2}\left\langle j,\alpha _j\right| \widehat{F%
}_A\otimes \widehat{I}_B\left| i,\alpha _i\right\rangle
=\sum\limits_iP_i\left\langle i\right| \widehat{F}_A\left| i\right\rangle .
\end{equation}
Therefore, we can see that the expectation value of a variable $\widehat{F}%
_A $ of the subsystem $A$ in mixed state can be converted into the
expectation value of a variable $\widehat{F}_A\otimes \widehat{I}_B$ of the
composite system with respect to a pure state. As a result, it is inevitable
to enlarge the original Hilbert space of the system to include the Hilbert
space of the surroundings, and hence the corresponding degree of freedom is
doubled.

Now, consider a thermodynamic system in thermal contact with a heat
reservoir. We know that a macroscopic system to be dealt with in statistical
physics consists of enormous number of microscopic elements interacting in a
complicated fashion with the surroundings, the energy intervals between
adjacent eigenstates will be much smaller than any disturbance due to the
surroundings. In other words, any small disturbance will cause transitions
between eigenstates, resulting in a uniform probabilistic distribution.
Consequently, the behavior of the dynamic system may be expected to be
described by certain probability laws$^8$. This implies that the state of a
system in equilibrium may be regarded as an incoherent superposition of
eigenstates of the system. As is well known, a collection of identical
systems used for studying probability characteristics is called an ensemble.
Therefore, the states of a thermodynamic system that are described by
statistical ensemble are mixed states, which are distinct from the pure
states which are described by state vectors in quantum mechanics$^9$. In
short, owing to the uncertain interaction between the thermodynamic system
and its surroundings, an uncertainty about micro-state is produced. In the
way, we know only an ensemble of possible states, say $|n>$, and the
probabilities $\rho _{n\text{ }}$of the system being in state $|n>$,
respectively. The information on the system is given by the density matrix.

In statistical physics we always deal with systems that interact with the
surroundings. Here we can regard the system plus its surroundings as a truly
isolated system. Following the former mark, we denote the thermodynamic
system and its surroundings by $A$ and $B$, respectively. Suppose $\widehat{F%
}_A$ is an observable of the thermodynamic system. As is well known that the
thermal statistical average of this observable in the thermal equilibrium
state is given by

\begin{equation}
\overline{\left\langle \widehat{F}_A\right\rangle }=Tr\left( \widehat{\rho }%
_A\widehat{F}_A\right) ,
\end{equation}
where, $\widehat{\rho }_A$ is the density operator of the thermodynamic
system; $Tr\left( \widehat{\rho }_A\widehat{F}_A\right) $denotes the trace
of the operator $\widehat{\rho }_A\widehat{F}_A$ and is the sum of all the
diagonal matrix elements of in any representation.

For definiteness, consider now a closed thermodynamic system with the
numbers of particles, volume and temperature fixed (The following discussion
can be easily generalized to the open system which may be described by grand
canonical ensemble.). Let $\widehat{H}_A$ be the Hamiltonian operator of the
system, then the density operator of the system under consideration has the
following form$^{10}$

\begin{equation}
\widehat{\rho }_A=\frac{e^{-\beta \widehat{H}_A}}{Z_A},
\end{equation}
where

\begin{equation}
Z_A=Tr\left( e^{-\beta \widehat{H}_A}\right)
\end{equation}
is the partition function of the system, and $\beta =\left( k_BT\right)
^{-1} $ , with $k_B$ is Boltzmann constant, $T$ the temperature of the
system.

Notice that the density operator and the Hamiltonian operator of the system
commute, they possess a complete orthonormal set of simultaneous
eigenstates. So, both $\widehat{\rho }_A$ and $\widehat{H}_A$ are diagonal
in energy representation. Assume we have the following eigenvalue equation
of the Hamiltonian

\begin{equation}
\widehat{H}_A\left| n\right\rangle =E_n\left| n\right\rangle .
\end{equation}
A set of linearly independent eigenvectors $\left| n\right\rangle $ of $%
\widehat{H}_A$ can be chosen to be orthogonal and normalized

\begin{equation}
\left\langle m|n\right\rangle =\delta _{mn},
\end{equation}
and the completeness relation can be express as

\begin{equation}
\sum\limits_n\left| n\right\rangle \left\langle n\right| =\widehat{I}_A.
\end{equation}
Then, the equation (17) becomes

\begin{equation}
\widehat{\rho }_A=\sum\limits_n\rho _n\left| n\right\rangle \left\langle
n\right| ,
\end{equation}
where

\begin{equation}
\rho _n=\frac{e^{-\beta E_n}}{Tre^{-\beta E_n}}
\end{equation}
is the probability for the system to have the energy eigenvalue $E_n$, and
it satisfies the normalization condition $\sum\limits_n\rho _n=1.$ Thus, the
thermal statistical average of the quantity $\widehat{F}_A$ can be expressed
as

\begin{equation}
\overline{\left\langle \widehat{F}_A\right\rangle }=\sum\limits_n\rho
_n\left\langle n\right| \widehat{F}_A\left| n\right\rangle .
\end{equation}

Suppose the statistical average of a quantity $\widehat{F}_A$ can be written
as expectation value with respect to a pure state $\left| O(\beta
)\right\rangle $. According to equation (14), we can write the following
relation

\begin{equation}
\overline{\left\langle \widehat{F}_A\right\rangle }=\left\langle O\left(
\beta \right) \left| \widehat{F}_A\otimes \widehat{I}_B\right| O\left( \beta
\right) \right\rangle .
\end{equation}
According to equations (2), the pure state $\left| O(\beta )\right\rangle $
can be expanded as

\begin{equation}
\left| O(\beta )\right\rangle =\sum\limits_{n,\mu }a_{n,\mu }\left| n,\mu
\right\rangle ,
\end{equation}
where $a_{n,\mu }=\left\langle n,\mu \right| \left| O(\beta )\right\rangle $%
, and the sets $\{|n>\}$ and $\{|\mu >\}$ are the complete orthonormal
eigenstates of the thermodynamic system and its surroundings, respectively.

In the energy representation, by means of Schmidt decomposition, $\left|
O(\beta )\right\rangle $ can be written as

\begin{equation}
\left| O(\beta )\right\rangle =\sum\limits_n\rho _n^{1/2}\left| n,\alpha
_n\right\rangle .
\end{equation}
with

\begin{equation}
\left| \alpha _n\right\rangle =\sum\limits_\mu \rho _n^{-1/2}a_{n,\mu
}\left| \mu \right\rangle .
\end{equation}
Obviously, the equation (25) automatically hold,

\begin{equation}
\sum\limits_n\rho _n\left\langle n\right| \widehat{F}_A\left| n\right\rangle
\equiv \sum\limits_{m,n}\rho _m^{1/2}\rho _n^{1/2}\left\langle m,\alpha
_m\right| \widehat{F}_A\otimes \widehat{I}_B\left| n,\alpha _n\right\rangle .
\end{equation}
Now, we denote $\alpha _n$ by $\widetilde{n}$, the pure state $\left|
O\left( \beta \right) \right\rangle $ can be written as

\begin{equation}
\left| O(\beta )\right\rangle =\sum\limits_n\rho _n^{1/2}\left| n,\widetilde{%
n}\right\rangle
\end{equation}
where $\rho _n$ is given by equation (23). Equation (30) is nothing but the
thermal vacuum state which was introduced in thermo field dynamic by
Takahashi and Umezawa$^1$. So, we can see clearly that the tilde system in
TFD is just the surroundings of the real system.

In addition, we can see from equation (30) that the thermal vacuum state $%
\left| O\left( \beta \right) \right\rangle $ is an infinite superposition of
the direct product state $\left| n\right\rangle \otimes \left| \widetilde{n}%
\right\rangle $. In general, the pure state $\left| O\left( \beta \right)
\right\rangle $ of the compound system cannot be expressed as a product $%
\left| n\right\rangle \otimes \left| \widetilde{n}\right\rangle $ of pure
states of its part (i. e., $A$ and $B$) and hence is an entangled state$%
^{11} $. We notice that, for the system under consideration, the pure state $%
\left| O\left( \beta \right) \right\rangle $ is a partly entangled pure
states, for the coefficients $\left( \rho _n\right) ^{1/2}$are not equal to
one another. Obviously, this conclusion is true for an open system which may
be described by grand canonical ensemble. In addition, for an isolated
system in statistical physics (In fact, it is not a truly isolated system.), 
$\rho _n$ in equation (30) becomes a constant. Since all of the expanding
coefficients in equation (30) are the same now, the state $\left| O\left(
\beta \right) \right\rangle $ corresponding to an isolated system is a
maximally entangled state$^{12}$.

\smallskip The density matrix of the compound system ($A$+$B$) corresponding
the state vector $\left| O\left( \beta \right) \right\rangle $ is given by

\begin{equation}
\widehat{\rho }_{AB}=\left| O\left( \beta \right) \right\rangle
_{ABAB}\left\langle O\left( \beta \right) \right| =\sum\limits_{m,n}\rho
_m^{1/2}\rho _n^{1/2}\left| m\right\rangle \otimes \left| \widetilde{m}%
\right\rangle \left\langle \widetilde{n}\right| \otimes \left\langle
n\right| .
\end{equation}
Because the $\left| O\left( \beta \right) \right\rangle $ is a pure state of
the compound system ($A$+$B$), the density matrix $\widehat{\rho }_{AB}$ is
non-diagonal. However, we are only interested in the real system $A$. When
we measure a dynamical variable of the real system, say energy, the state of
the system will transfer to one of its energy eigenstate and lead to the
quantum decoherence$^{13}$. Mathematically, this means the density matrix
will be traced in the tilde states. That is we should describe the real
system by means of the following reduced density matrix$^{5,6}$

\begin{equation}
\widehat{\rho }_A=Tr_B\left( \widehat{\rho }_{AB}\right)
=\sum\limits_l\left\langle \widetilde{l}\right| \widehat{\rho }_{AB}\left| 
\widetilde{l}\right\rangle =\sum\limits_{l,m,n}\rho _m^{1/2}\rho
_n^{1/2}\left\langle \widetilde{l}\right| \left| m\right\rangle \otimes
\left| \widetilde{m}\right\rangle \left\langle \widetilde{n}\right| \otimes
\left\langle n\right| \left| \widetilde{l}\right\rangle =\sum\limits_n\rho
_n\left| n\right\rangle \left\langle n\right| ,
\end{equation}
which is the same as the corresponding one in statistical physics (see
equation (22) ). So, the surroundings forms an pure entangled states with
the thermodynamic system upon tracing out the former we can get the correct
mixed state for the latter system.

In summary, by the unavoidable interaction of the thermodynamic system in
question with the surroundings, the thermodynamic system will be in a mixed
state. If the thermal statistical average is replaced by a vacuum
expectation value in a pure state, then the Hilbert space of the
thermodynamic system should be generalized to include the one of the
surroundings, and the corresponding operator become a direct product which
include the operator of the system and the identity operator of the
surroundings.

\noindent {\LARGE Acknowledgments}

This project is supported by the Chinese Education Foundation(grant no.
1999014105).


\begin{references}
\bibitem{}  Y. Takahashi and H. Umezawa, {\sl CollectivePhenomena} {\bf 2}
55 (1975).

\bibitem{}  H. Umezawa, H. Matsumoto and M. Tachiki, {\it Thermo Field
Dynamics and Condensed States}, (North-Holland, Amsterdam, 1982).

\bibitem{}  Y. X. Gui, {\sl Phys. Rev. D} {\bf 46} 1869 (1992).

\bibitem{}  {\it Proceedings of the 3rd Workshop on Thermal Field Theories
and Their Applications}, edited by F. C. Khanna, et al, (World Scientific
Publishing Co., Singapore, 1994).

\bibitem{}  J. von Neumann, {\it Mathematical Foundations of Quantum
Mechanics}, (Princeton University Press, Princeton, 1955).

\bibitem{}  Landau, L. D., {\sl Zeit. Phys.} {\bf 45} 430 (1927).

\bibitem{}  A. Peres, {\it Quantum Theory: Concepts and Methods} (Kluwer,
Dordrecht, 1993).

\bibitem{}  M. Toda, R. Kubo, N. Sait\^{o}S, {\it Statistical Physics} {\bf I%
}, (2nd ed, Springer, Berlin, Heidelberg, 1997).

\bibitem{}  D. Landau, E. M. Lifshiz, S{\it tatistical Mechanics},
(Pergamon, Oxford, 1958).

\bibitem{}  Kerson Huang, {\it Statistical Mechanics}, (2nd ed, John Wiley
\& Sons, New York, 1987).

\bibitem{}  E. Schr\"{o}dinger, {\sl Proc. Cambridge. Philos. Soc.} {\bf 31}
555 (1935).

\bibitem{}  S. L. Braunstein, A. Mann and M. Revzen, {\sl Phys. Rev. Lett.} 
{\bf 68} 3259 (1992).

\bibitem{}  W. H. Zurek, {\sl Physics Today} {\bf 44} 36 (1991).
\end{references}
\end{document}